\documentclass[twocolumn,natbib]{svjour3}

\smartqed  								
\usepackage{mathptmx}				    
\usepackage{graphicx}

\usepackage{fixltx2e}					

\usepackage{amsmath,amscd,amssymb}
\usepackage[misc]{ifsym}				
\usepackage[usenames,dvipsnames]{color}	

\usepackage[absolute]{textpos}

\newcommand{\todo}[1]{\textcolor{blue}{#1}}

\newcommand{\revision}[1]{\textcolor{blue}{#1}}

\usepackage{ifthen}
\newcommand{\switch}[1]{\ifthenelse{\equal{#1}{0}}{
	\renewcommand{\todo}[1]{}
	\renewcommand{\revision}[1]{##1}
	}{}}
\switch{0}		

\newcommand{\matstyle}[1]{#1}
\newcommand{\greekmatstyle}[1]{#1}
\newcommand{\Psimat}{\greekmatstyle{\Psi}}
\newcommand{\Phimat}{\greekmatstyle{\Phi}}
\newcommand{\Fmat}{\matstyle{F}}
\newcommand{\Fhatmat}{\hat{\Fmat}}
\newcommand{\Gmat}{\matstyle{G}}
\newcommand{\Mmat}{\matstyle{M}}
\newcommand{\Umat}{\matstyle{U}}
\newcommand{\Amat}{\matstyle{A}}
\newcommand{\Bmat}{\matstyle{B}}
\newcommand{\Ahatmat}{\hat{\Amat}}

\newcommand{\vecstyle}[1]{{#1}}
\newcommand{\fvec}{\vecstyle{f}}
\newcommand{\fhatvec}{\hat{\fvec}}
\newcommand{\gvec}{\vecstyle{g}}
\newcommand{\zerovec}{\vecstyle{0}}
\newcommand{\wvec}{\vecstyle{w}}
\newcommand{\vvec}{\vecstyle{v}}

\journalname{Experiments in Fluids}

\usepackage{hyperref}			
\usepackage[all]{hypcap}		
\hypersetup{colorlinks=true,	
	linkcolor=red,			    
	citecolor=blue,		        
    urlcolor=cyan               
	}

\title{Toward compressed DMD: spectral analysis of fluid flows using sub-Nyquist-rate PIV data}

\author{Jonathan H. Tu
\and Clarence W. Rowley
\and J. Nathan Kutz
\and Jessica K. Shang}

\institute{J. H. Tu ( {\Letter} ), C. W. Rowley, J. K. Shang
\at Department of Mechanical and Aerospace Engineering \\
Princeton University, Princeton, NJ 08544, USA \\
\email{jhtu@berkeley.edu} \\
\and J. Nathan Kutz
\at Department of Applied Mathematics \\
University of Washington, Seattle, WA 98195, USA
}

\date{Received: \todo{X} / Revised: \todo{X} / Accepted: \todo{X} / Published online: \todo{X}} 

\begin{document}

\maketitle


\begin{textblock*}{6in}(0.586in,0.6in)
    \noindent
    \texttt{Manuscript submitted to Experiments in Fluids.}
\end{textblock*}


\begin{abstract}
    Dynamic mode decomposition (DMD) is a powerful and increasingly popular tool for performing spectral analysis of fluid flows.
    However, it requires data that satisfy the Nyquist-Shannon sampling criterion.
    In many fluid flow experiments, such data are impossible to capture.
    We propose a new approach that combines ideas from DMD and compressed sensing.
    Given a vector-valued signal, we take measurements randomly in time (at a sub-Nyquist rate) and project the data onto a low-dimensional subspace.
    We then use compressed sensing to identify the dominant frequencies in the signal and their corresponding modes.
    We demonstrate this method using two examples, analyzing both an artificially constructed test dataset and particle image velocimetry data collected from the flow past a cylinder.
    In each case, our method correctly identifies the characteristic frequencies and oscillatory modes dominating the signal, proving the proposed method to be a capable tool for spectral analysis using sub-Nyquist-rate sampling.
\end{abstract}



\section{Introduction}
\label{sec:intro}

Many dynamical systems exhibit oscillatory behavior; fluid mechanical systems are no exception.
In recent years, many have turned to dynamic mode decomposition (DMD)~\citep{rowleyJFM09,schmidJFM10,tuArxiv13} as a useful tool for analyzing such systems.
Not only can DMD identify characteristic flow frequencies, but the corresponding modes may elucidate features of the underlying fluid mechanics.
Unfortunately, DMD requires data that satisfy the Nyquist-Shannon sampling criterion~\citep{nyquistTAIEE28,shannonPIRE49}, which may not always be available in practice, due to sensor limitations.
For example, in fluid flow experiments DMD is typically applied to particle image velocimetry (PIV) data.
While time-resolved PIV systems are presently capable of sampling rates as high as 800~Hz, such systems are extremely expensive and thus rare; standard PIV equipment is limited to sampling frequencies on the order of 15~Hz~\citep{tuEF13}.

In the signal processing community, there has been a growing emphasis on dealing with time resolution issues using a method called \emph{compressed sensing}~\citep{donohoIEEETIT06b,candesIEEETIT06_1,candesIEEETIT06_2}.\footnote{Compressed sensing is also known as ``compressive sampling'' or ``compressive sensing'' and is closely related to ``sparse approximation''/``sparse reconstruction''/``sparse recovery'' methods.}
Compressed sensing relies on the fact that many signals of interest are sparse in frequency space.
If we sample such signals randomly in time, then we can reconstruct them accurately using $\ell_1$ minimization techniques or greedy algorithms, even if the samples are taken at a sub-Nyquist rate.
(For a review of compressed sensing theory, we refer the reader to~\cite{baraniukIEEESPM07},~\cite{candesIEEESPM08}, and~\cite{bryanSIAMR13}.)
This approach has proven successful in a number of applications, including dynamic MRI~\citep{lustigMRM07,gamperMRM08}, facial recognition~\citep{wrightIEEETPAMI09}, imaging~\citep{duarteIEEESPM08,rombergIEEESPM08}, radar~\citep{hermanIEEETSP09,potterPIEEE10}, classification tasks~\citep{bruntonbArxiv13,bruntonsArxiv13}, and reconstruction of turbulent flow fields~\citep{baiAIAA13}.

While typically applied to scalar-valued signals, compressed sensing algorithms extend readily to vector-valued signals.
As such, in theory these methods can be applied directly to PIV data collected from fluids experiments.
However, in reality this is usually not feasible; the fine spatial resolution needed to accurately resolve pertinent flow features makes PIV data too large for standard compressed sensing algorithms.
But despite their frequent representation as high-dimensional vectors, many fluid flows actually evolve in a low-dimensional subspace.
Projecting PIV data onto this subspace, we can obtain a low-dimensional encoding that makes compressed sensing tractable.

We propose a method for computing temporally oscillating modes from sub-Nyquist-rate PIV data, combining concepts from DMD and compressed sensing.
DMD is closely related to proper orthogonal decomposition (POD): the DMD modes are linear combinations of POD modes and the DMD eigenvalues come from the POD projection of an approximating linear operator~\citep{schmidJFM10,tuArxiv13}.
Here, we use POD projections to represent high-dimensional PIV data using low-dimensional vectors, in order to reduce computational costs.
We then perform compressed sensing on these vectors of POD coefficients, lifting the resulting modes to the original space by taking linear combinations of POD modes, just as in DMD.
Not only does this method rely on POD in the same way that DMD does, but POD bases are also optimal for reconstructing datasets.

We note that compressed sensing methods have been used in conjunction with DMD previously, in work by~\cite{jovanovicArxiv13} and~\cite{bruntonsArxiv13b}.
However, their approaches differ from that taken here, in which sparsity is leveraged to overcome time-resolution issues.
\cite{jovanovicArxiv13} enforce sparsity as a post-processing step designed to select modes of interest.
Those modes are computed using a standard DMD computation, relying on time-resolved data.
\cite{bruntonsArxiv13b} use compressed sensing to more efficiently perform DMD computations, subsampling in space but again making use of time-resolved data.

We demonstrate our method through two extended examples.
In the first, we construct a canonical dataset in which we superpose two Gaussian spatial fields oscillating in time at different frequencies.
We add noise to the signal to test the robustness of our method.
By construction, the signal is almost exactly sparse (it is heavily dominated by two frequencies), so a compressed sensing approach is reasonable.
However, we choose the amplitudes of these Gaussian fields such that the resulting POD modes mix the oscillatory structures together.
In the second example, we apply our method to experimental PIV data collected from the flow past a cylinder.
This flow is dominated by a single frequency (the cylinder shedding frequency), but the data are not precisely sparse, only approximately so.
Each of these examples poses different challenges for our method, but in both cases we successfully identify the correct frequencies and modes.

The rest of this paper is structured as follows: Sec.~\ref{sec:methods} details our numerical method, first introducing the fundamental concepts of compressed sensing theory and then describing how we implement those ideas in practice.
Then in Sec.~\ref{sec:results}, we discuss results from the two example applications described above.
Finally, we summarize and offer directions for future work in Sec.~\ref{sec:conclusions}.


\section{Numerical method}
\label{sec:methods}

In this section we introduce the basic concepts of compressed sensing and describe how we apply those concepts to compute temporally oscillating spatial modes from sub-Nyquist-rate data.
First, we describe how compressed sensing can be used to reconstruct scalar signals.
Then we show how it can be extended to vector-valued signals.
For high-dimensional vectors, the computational costs of standard compressed sensing methods can be prohibitive; we propose the use of POD to reduce the dimension of the problem, lowering those costs.
Finally, we describe strategies for collecting data samples suitable for compressed sensing and then summarize our numerical method.

\subsection{Scalar signals}

The field of compressed sensing has undergone astounding growth since the foundational works by \citet{donohoIEEETIT06b}, \citet{candesIEEETIT06_1}, and \citet{candesIEEETIT06_2} were published in 2006.
We review the key concepts here.
(For more a more in-depth introduction to these topics, we again refer the reader to~\cite{baraniukIEEESPM07},~\cite{candesIEEESPM08}, and~\cite{bryanSIAMR13}.)
Consider a signal $\fvec \in \mathbb{R}^n$.
For instance, $\fvec$ could consist of $n$ sequential measurements taken from a hot wire velocity probe.
We assume that these measurements are taken at a rate such that $\fvec$ captures all dynamics of interest.

Typically, $\fvec$ will not be sparse in the standard basis for $\mathbb{R}^n$ (comprising the vectors $(1,0,0,\ldots)$, $(0,1,0,\ldots)$, and so on).
That is, a large number of these basis vectors are required to accurately describe $\fvec$.
We say that $\fvec$ is \emph{compressible} if there exists a basis $\Psimat$ such that the representation of $\fvec$ in $\Psimat$ is approximately sparse.
Specifically, we say that $\fvec$ is $k$-sparse in the basis $\Psimat$ if
\begin{equation}
    \label{eq:cs_f_as_Psi}
    \fvec = \Psimat \fhatvec,
\end{equation}
where $\Psimat \in \mathbb{R}^{n \times n}$ and $\fhatvec \in \mathbb{R}^n$, with $\fhatvec$ having only $k \ll n$ nonzero values.
(The less precise descriptor ``compressible'' requires only that $\fhatvec$ have few large coefficients relative to the number of small ones.)
The potential for savings is clear: rather than storing $n$ values to describe the signal $\fvec$, we can get away with storing only the $k$ nonzero elements of $\fhatvec$.
This is the principle on which JPEG-2000 compression is built~\citep{taubmanJPEG2000ICFSP01,candesIEEESPM08}.

Now suppose that we do not have access to the full signal~$\fvec$.
Instead, all we know is an $m$-dimensional linear measurement
\begin{equation}
    \label{eq:g_as_f}
    \gvec = \Phimat^T \fvec,
\end{equation}
where $\Phimat$ is an $n \times m$ matrix.
We can think of the columns of $\Phimat$ as waveforms that we use to measure $\fvec$.
For instance, if $\Phimat$ contains sinusoids, then $\gvec$ contains Fourier coefficients.

We are interested in the case where $m \ll n$, i.e., the undersampled case, for which Eq.~\eqref{eq:g_as_f} is underdetermined.
As such, we cannot solve for $\fvec$ from a knowledge of $\gvec$; the solution, if it exists, is not unique.
But suppose we substitute for $\fvec$ using Eq.~\eqref{eq:cs_f_as_Psi}, giving us
\begin{equation}
    \label{eq:g_as_fhat}
    \gvec = \Phimat^T \Psimat \fhatvec.
\end{equation}
Though $\fhatvec$ is also an $n$-dimensional vector, it only has $k$ nonzero elements, where we assume that $k < m \ll n$.
A standard approach in compressed sensing is to determine $\fhatvec$ by solving the following optimization problem:
\begin{equation}
    \label{eq:cs_opt_problem}
    \min_{\fhatvec \in \mathbb{R}^n} \big\| \fhatvec \big\|_1 \quad \text{subject to} \quad \gvec = \Phimat^T \Psimat \fhatvec.
\end{equation}
This can be interpreted as follows: of all vectors $\fhatvec$ that are consistent with our measurement $\gvec$, we are interested in finding the one with the smallest $\ell_1$ norm.

We choose the $\ell_1$ norm because it promotes sparsity and allows Eq.~\eqref{eq:cs_opt_problem} to be solved using a linear program.
Compared to the more common $\ell_2$ norm, used for instance in solving least-squares problems, the $\ell_1$ norm more harshly penalizes small nonzero elements in $\fhatvec$, which we know to be a sparse vector.
In theory we would like to minimize the cardinality\footnote{Though technically not a norm, the cardinality of a vector is often referred to as its $\ell_0$ norm.} of $\fhatvec$ (the number of nonzero components), but that minimization problem is NP-complete and numerically unstable~\citep{baraniukIEEESPM07}.
As such, we use the $\ell_1$ norm as a computationally tractable proxy.

It was shown by~\cite{donohoIEEETIT06b} and~\cite{candesIEEETIT06_1} that in some cases, solving Eq.~\eqref{eq:cs_opt_problem} can recover $\fhatvec$ exactly if $\fhatvec$ is $k$-sparse, or very accurately if $\fhatvec$ is compressible.
Much of the compressed sensing literature deals with finding conditions on $\Phimat$ and $\Psimat$ for which these results hold.
For instance, the columns of $\Phimat$ and $\Psimat$ should be chosen to be maximally \emph{incoherent}.
Many proofs also rely on $\Phimat^T \Psimat$ obeying the \emph{restricted isometry property}~\citep{dickIEEESFPCCM00,candesIEEETIT05,candesCRASP08}.
These topics are outside the scope of this discussion and furthermore are most applicable to situations in which we have freedom to choose the measurement matrix.
In this work, we restrict ourselves to the case that $\Psimat$ describes a Fourier basis and that the columns of $\Phimat$ compose a subset of the standard basis (see Sec.~\ref{ssec:basis_choice} for more details).
A more relevant theoretical result is that solving Eq.~\eqref{eq:cs_opt_problem} yields the best $k$-sparse approximation to $\fvec$ even if $\fvec$ is not exactly $k$-sparse (e.g., if it is only compressible)~\citep{candesIEEESPM08}.
Furthermore, this procedure is robust to measurement noise~\citep{candesIEEESPM08}.

Closely related to compressed sensing is the field of \emph{sparse approximation}.
Just as in compressed sensing, the goal of sparse approximation methods is to find the best sparse representation of a $k$-sparse or compressible signal.
However, rather than solving an $\ell_1$ minimization problem, sparse approximation methods make use of greedy algorithms.
These algorithms are iterative: upon each iteration they add another basis vector (column of $\Psimat$) to the support of $\fhatvec$.
By construction, the resulting estimate of $\fhatvec$ will be sparse, as $j$ iterations will yield $j$ nonzero basis coefficients; the rest are assumed to be zero.
There are many different greedy algorithms used for sparse approximation.
In this work, we deal only with orthogonal matching pursuit (OMP), due to its simplicity~\citep{troppIEEETIT04,troppIEEETIT07}.
(CoSaMP is a similar algorithm that is also popular~\citep{needellACHA09}.)
The theoretical guarantees on OMP are similar to those for compressed sensing: under certain technical conditions (again outside the scope of this work), OMP exactly reproduces $k$-sparse vectors and closely approximates compressible ones~\citep{troppIEEETIT04}.

\subsection{Choice of basis, measurement}
\label{ssec:basis_choice}

Much of the theoretical research on compressed sensing deals with characterizing matrices $\Phimat$ and $\Psimat$ for which the method will succeed.
In this work, we are motivated by practical concerns, and as such are restricted in our choices of $\Phimat$ and $\Psimat$.
Because we are concerned with temporally oscillatory behavior, we choose $\Psimat$ such that $\fhatvec$ contains Fourier coefficients, assuming sparsity in the Fourier basis.
From~\eqref{eq:cs_f_as_Psi}, we see that this means $\Psimat$ is the matrix representation of the inverse discrete Fourier transform (DFT).
Our use of a Fourier basis is consistent with DMD, which is typically used to decompose a dataset into spatial modes that each oscillate at a fixed temporal frequency.
(However, in DMD, the modes may also exponentially grow or decay in time, whereas our Fourier description is purely oscillatory.)

For ease of implementation, we assume that our measurement $\gvec$ simply corresponds to values of $\fvec$ sampled at particular instants in time.
Suppose that $\fvec$ corresponds to a fast hot wire probe signal.
The first element of $\fvec$ is the value of the probe signal at time $t=0$.
The second element is the value at $t = \Delta t$, the third element corresponds to $t = 2 \Delta t$, and so on.
Now suppose that for our measurement $\gvec$, we sample our probe signal at $t=0$.
Then the first column of $\Phimat$ is $(1,0,0,0,\ldots)^T$.
If we wait until $t = 2 \Delta t$ to get our next sample, then the second column of $\Phimat$ is $(0,0,1,0,\ldots)^T$.
Thus we see that the columns of $\Phimat$ compose a subset of the standard basis: each column contains only zeroes, except for one entry with value $1$.
We can think of our measurement waveforms as Dirac delta functions.
As it turns out, delta functions and sinusoids are maximally incoherent, an important property for compressed sensing to work~\citep{candesIEEESPM08}.

\subsection{Vector-valued signals}

In this work we are concerned with vector-valued signals $\Fmat \in \mathbb{R}^{n \times p}$; the compressed sensing literature refers to such signals as ``multiple-measurement vectors''~\citep{cotterIEEETSP05,malioutovIEEETSP05,chenIEEETSP06,troppSP06a,troppSP06b,eldarIEEETIT09}.
As before, $n$ corresponds to the number of temporal measurements and $p$ is the number of values measured at a given instant in time.
If $\Fmat$ corresponds to a rake of hot wire sensors, then $p$ is the number of hot wires.
If $\Fmat$ corresponds to PIV velocity fields, then each field is reshaped into a row vector and $p$ is the number of grid points in the velocity field multiplied by the number of velocity components measured (typically two).
In this case, we observe that rows (of $\Fmat$) correspond to points in time and columns to points in space.

We assume that there exists a basis $\Psimat$ in which the representation of $\Fmat$ is sparse.
Since $\Fmat$ is a matrix, we must be careful in defining what we mean by sparse.
For a vector-valued signal, we rewrite Eq.~\eqref{eq:cs_f_as_Psi} as
\begin{equation}
    \label{eq:F_as_Psi}
    \Fmat = \Psimat \Fhatmat,
\end{equation}
where $\Fhatmat \in \mathbb{R}^{n \times p}$.
In the simple case that $p=1$, for which Eq.~\eqref{eq:F_as_Psi} reduces to Eq.~\eqref{eq:cs_f_as_Psi}, sparsity requires that $\fvec$ have few large elements.
When $p > 1$, the elements of $\fvec$ correspond to rows of $\Fmat$, so we require that there be few rows of $\Fmat$ with large norm.
Letting $\Gmat$ be a vector-valued measurement analagous to $\gvec$, we can rewrite Eq.~\eqref{eq:cs_opt_problem} as
\begin{equation}
    \label{eq:cs_mat_opt_problem}
    \min_{\Fhatmat \in \mathbb{R}^{n \times p}} \big\| \Fhatmat \big\|_{1,q} \quad \text{subject to} \quad \Gmat = \Phimat^T \Psimat \Fhatmat,
\end{equation}
where $\Gmat \in \mathbb{R}^{m \times p}$ and the mixed norm $\|\cdot\|_{1,q}$ of a matrix $\Mmat$ is defined as
\begin{equation}
    \label{eq:def_mixed_norm}
    \|\Mmat\|_{1,q} \triangleq \sum_{i=0}^{n-1} \left| \left(\sum_{j=0}^{p-1} |\Mmat_{i,j}|^q \right)^{1/q}  \right|.
\end{equation}

This norm can be interpreted as taking the $\ell_q$ norm of each row, stacking these values in a vector, and then taking the $\ell_1$ norm of the vector of $\ell_q$ norms.
The choice of $q$ weights the relative importance of nonzero entries that occur in the same row versus those that occur in different ones.
For instance, if $q=1$, then we have an $\ell_1$ equivalent of the Frobenius norm for matrices and all nonzero elements are penalized equally.
However, in some applications we may expect that only a few rows of $\Fhatmat$ will contain nontrivial entries, but within those rows we may have no expectation of sparsity.
In this case we would choose $q > 1$ to decrease the penalty on nonzero elements within rows.
(For other examples demonstrating the use of compressed sensing with mixed norms, see~\cite{cotterIEEETSP05,malioutovIEEETSP05,chenIEEETSP06,troppSP06a,troppSP06b,eldarIEEETIT09}.)
Recall from Sec.~\ref{ssec:basis_choice} that in this work we choose $\Psimat$ to be the DFT basis.
Then each row of $\Fhatmat$ corresponds to a particular frequency.
Our notion of row sparsity is then natural, as it corresponds to a signal dominated by a small number of frequencies.

We note that one could theoretically perform compressed sensing on the columns of $\Fmat$ individually, treating each as a scalar signal.
Each computation would yield a sparse coefficient vector $\fhatvec$.
However, there would be no guarantee that the sparse elements would occur in the same entries across computations.
For a Fourier basis, that means that while each computation would identify a small number of dominant frequencies, these frequencies might vary from computation to computation.
This highlights an advantage of the vector-valued approach: sparsity is enforced using all of the data simultaneously.

\subsection{Efficiency through POD projection}
\label{ssec:pod-projection}

In practice, solving the optimization problem given by Eq.~\eqref{eq:cs_mat_opt_problem} can be computationally prohibitive when the matrix $\Fmat$ is large.
For PIV velocity fields, the dimension $p$ corresponds to the number of grid points multiplied by the number of velocity components, which can easily exceed $10^5$.
Fortunately, many fluid flows evolve in relatively low-dimensional subspaces.
We can take advantage of this to make compressed sensing feasible for PIV data.

Consider a vector-valued signal $\Fmat$ where each row corresponds to a PIV velocity field.
(We reshape each velocity field into a row vector and concatenate each velocity component to get a single vector describing the entire flow field.)
The transposed matrix $\Fmat^T$ is often referred to as a ``snapshot'' matrix in the DMD and POD literature, as each of its columns describes a snapshot of the flow field at an instant in time.\footnote{We note the convention in fluid mechanics is that each \emph{column} of the snapshot matrix corresponds to an instant in time.
For compressed sensing, the convention is reversed: each \emph{row} of the signal matrix corresponds to a particular instant.}
For large $p$, we can compute the POD modes of $\Fmat^T$ efficiently using the method of snapshots~\citep{sirovichQAM87_2,rowleyIJBC05}.
The projection of $\Fmat^T$ onto the first $r$ POD modes is given by
\begin{equation}
    \label{eq:Fhat_as_POD}
    \mathbb{P}_r\Fmat^T = \Umat_r \Umat_r^T \Fmat^T,
\end{equation}
where $\Umat_r \in \mathbb{R}^{p \times r}$ is a matrix whose columns are POD modes.
We refer to the $r \times n$ matrix
\begin{equation}
    \label{eq:def_A}
    \Amat_r^T \triangleq \Umat_r^T \Fmat^T
\end{equation}
as the matrix of POD coefficients.\footnote{Again, we use the compressed sensing convention: \emph{rows} of $\Amat_r$ correspond to instants in time.}
We can project a vector-valued measurement $\Gmat$ in the same way, yielding the POD coefficient matrix
\begin{equation}
    \label{eq:def_B}
    \Bmat_r^T = \Umat_r^T \Gmat^T,
\end{equation}
where $\Bmat_r \in \mathbb{R}^{m \times r}$.
(Recall, $n$ is the number of time points, $m$ is the number of samples in time, $p$ is the size of the data vector, e.g., the number of grid points, and $r$ is the number of POD modes.)

Since $\Fmat$ has a sparse representation in $\Psimat$, so should $\Amat_r$; $\Amat_r$ describes the same behavior in a different coordinate system.
Then we can write
\begin{equation}
    \label{eq:def_Ahat}
    \Amat_r = \Psimat \Ahatmat_r,
\end{equation}
where $\Ahatmat_r \in \mathbb{R}^{n \times r}$, and we can apply compressed sensing to $\Amat_r$ by solving
\begin{equation}
    \label{eq:cs_mat_opt_problem_POD}
    \min_{\Ahatmat_r \in \mathbb{R}^{n \times r}} \big\| \Ahatmat_r \big\|_{1,q} \quad \text{subject to} \quad \Bmat_r = \Phimat^T \Psimat \Ahatmat_r.
\end{equation}
By using the mixed norm $\|\cdot\|_{1,q}$, we enforce row sparsity, meaning that only a few rows of $\Ahatmat_r$ should contain nontrivial values.
We emphasize that in this work sparsity is assumed in the Fourier basis and \emph{not} in the POD basis; the latter is used only as a means to reduce computational costs.
This is in contrast to recent work by~\cite{baiAIAA13} and~\cite{bruntonsArxiv13}, in which sparsity is achieved via projection onto a POD basis.

Recall that we are interested in computing spatial modes that each oscillate in time with a fixed frequency, similar to DMD.
We can find such modes by using $\Ahatmat_r$ to linearly combine the POD modes.
The matrix $\Ahatmat_r$ has rows that each correspond to a frequency and columns that each correspond to a POD mode.
Then each column of the product $\Umat_r \Ahatmat^T$ is a spatial field corresponding to a particular frequency.
These are the equivalent of DFT modes, computed using sub-Nyquist-rate data.

In an abstract way, this method is quite similar to DMD.
Recall that DMD is closely related to POD, with the DMD modes computed as a linear combination of POD modes~\citep{schmidJFM10,tuArxiv13}.
The rest of the DMD procedure can be considered a computation to determine the proper coefficients for this linear combination.
The result is a set of modes that each correspond to a particular frequency (and growth/decay rate).
Similarly, by construction the columns of $\Umat_r \Ahatmat_r^T$ are linear combinations of POD modes.
We can consider the compressed sensing procedure as a computation to determine the right coefficients for this linear combination.
However, unlike DMD, because we assume that $\Psimat$ is a DFT basis, the compressed sensing modes are purely oscillatory; there are no growth rates.

\subsection{Sampling strategy}
\label{ssec:sampling-strategy}

The allure of compressed sensing is that it can be used to circumvent the Nyquist-Shannon sampling criterion.
A key requirement is that the signal of interest must be compressible, but this is not uncommon; many signals are dominated by a few characteristic frequencies.
The other unique aspect of compressed sensing is that it relies on random sampling strategies.
Results on whether or not a compressible signal can be recovered generally focus on the number of measurements required, with no constraints on the sampling other than that it is random.
(For decision-making purposes based on spatial measurements,~\cite{bruntonbArxiv13} provide a method for finding optimal sensor locations, showing that well-designed sampling strategies can outperform random sampling.)
However, not all random sampling strategies will work.
For instance, if we happen to sample a scalar signal at only zero crossings, then we have $\gvec = \zerovec$, and there is obviously not enough information to reconstruct a nontrivial signal.
Furthermore, in practice a truly random sampling may not be possible due to physical constraints; the minimum time between samples is a common limitation.

In this work we develop sampling strategies based on physical intuition.
We assume the elements of the nominal signal $\fvec$ correspond to times $t=0$, $t=\Delta t$, $t=2 \Delta t$, and so on, with $\Delta t$ small enough that $\fvec$ captures all dynamics of interest.
(For simplicity we refer to scalar signals in this discussion, though everything extends to vector-valued signals.)
Motivated by applications to PIV data, we assume that the closest any two samples can be in time is $s_\text{min} \Delta t$.
We know from the Nyquist-Shannon sampling criterion that if we sample $\fvec$ at a fixed rate corresponding to $s_\text{min} \Delta t$, we may alias the signal and be unable to recover any oscillations with frequencies faster than $1 / (2 s_\text{min} \Delta t)$.
Thus, we do not expect that any (even if random) subset of those samples will suffice for compressed sensing.

Instead, we assume that though we have a minimum separation between samples, we have enough accuracy to sample any element of $\fvec$, so long as it is not within $s_\text{min}$ samples of the previous one.
That is, we are not interested in the fastest possible uniform sampling, which is given by data collected at times $t=0$, $t=s_\text{min} \Delta t$, $t=2 s_\text{min} \Delta t$, and so on.
Rather, we make use of the fact that we can collect data at $t = t^*$ and $t=t^*+(s_\text{min} + j)\Delta t$ for any $j$.
Intuitively, this allows us to sample all phases of our signal, even though we cannot do so in a frequency-resolved manner.
Applying this strategy in a random fashion (letting $j$ vary randomly), the sampled signal should contain as much information as a truly random sampling, as is usually considered in compressed sensing.

Based on this intuition, we propose the following sampling strategy, which we refer to as the ``minimum/maximum separation strategy'':
\begin{enumerate}
    \item Define a minimum separation between samples $s_\text{min}$.
    \item Define a maximum separation between samples $s_\text{max}$.
    \item Sample the signal $\fvec$ such that the time between samples is given by $j \Delta t$, where $j$ is random and uniformly distributed between $s_\text{min}$ and $s_\text{max}$.
\end{enumerate}
Alternatively, one could imagine randomly perturbing an otherwise regular sampling rate.
The time between samples would then be given by $(s_\text{avg} + j)\Delta t$, where $j$ is random and uniformly distributed between $-s_\text{pert}$ and $s_\text{pert}$, and $s_\text{pert}$ is the maximum allowable perturbation in the sample separation.
However, this is just a special case of the minimum/maximum separation strategy, with $s_\text{avg} = (s_\text{min}+s_\text{max})/2$ and $s_\text{pert} = (s_\text{max}-s_\text{min})/2$.

The value of $s_\text{min}$ can be chosen such that no samples are collected faster than allowed by the maximum sampling rate.
Intuitively, $s_\text{max}$ should be large enough that all phases of the signal are sampled, in order to avoid aliasing.
One rule of thumb is to make sure that the maximum spacing between samples is at least as large as $1/f_\text{min}$, where $f_\text{min}$ is a characteristic slow frequency.

\subsection{Summary of method}

We summarize the steps of our method here.
\begin{enumerate}
    \item Define a strategy for sampling randomly in time (see Sec.~\ref{ssec:sampling-strategy}).
    \item Use this strategy to generate a ``chirp signal'' of 1's and 0's, where a 1 corresponds to a time when a sample should be collected.
    When the chirp signal has value 0, no data should be collected.
    \item Set up a triggering system such that data are only collected when the value of the chirp signal is 1.
    \item Collect data according to the chirp signal.
    \item Compute POD modes from the data.
    \item Choose a set of $r$ POD modes to represent the data, for instance setting a threshold for the amount of energy captured by the modes.
    This defines the matrix $\Umat_r$.
    \item Project the data onto the POD modes, resulting in a matrix of sampled POD coefficients $\Bmat_r$.
    \item Solve the optimization problem given by Eq.~\eqref{eq:cs_mat_opt_problem_POD}, where $n$ is determined by the time elapsed between the first and last data samples.
    \item Compute the compressed sensing modes as the columns of $\Umat_r \Ahatmat_r^T$.
\end{enumerate}
We note that for especially long signals (large $n$), the optimization problem given by Eq.~\eqref{eq:cs_mat_opt_problem_POD} can be replaced with a greedy algorithm such as OMP.
In that case, only the nontrivial rows of $\Ahatmat_r$ will be computed, but the computation of the compressed sensing modes as a linear combination of POD modes is unchanged.


\section{Results}
\label{sec:results}

In this section we present two extended examples that demonstrate the capabilities of the method described above.
The first deals with a numerical dataset that we construct, designed to test various features of our method.
The second applies our method to data collected from a fluid flow experiment.
In both cases, we are able to correctly identify the characteristic frequencies and oscillatory modes that dominate the signal of interest, using only sub-Nyquist-rate samples.

\subsection{Canonical dataset}

\begin{figure}[b]
    \centering
    \includegraphics{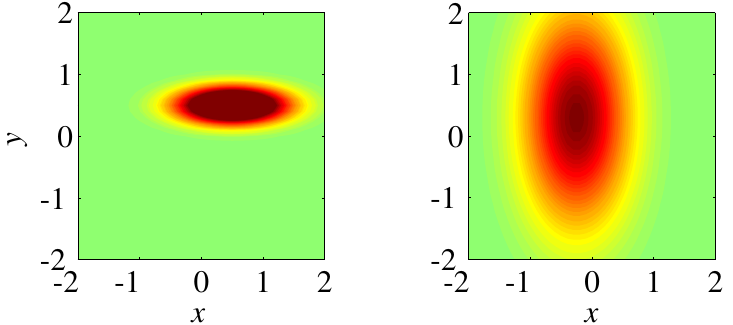}
    \caption{True oscillatory modes for the canonical dataset, as defined in Eq.~\eqref{eq:toy_ex_true_modes}.
    (Left) $\vvec_1$; (right) $\vvec_2$.}
    \label{fig:toy_ex_true_modes}
\end{figure}

The vast literature on compressed sensing leaves very little doubt that $\ell_1$ minimization and greedy algorithms can in fact reconstruct compressible signals.
Thus the features of our method that require verification are the sampling strategy and the use of a POD projection to reduce computational costs.
As a test, we consider a dataset of the form
\begin{equation}
    \label{eq:canon_signal}
    \fvec(t) = \sin(\omega_1 t) \vvec_1 + \sin(\omega_2 t) \vvec_2 + 0.1 \wvec(t).
\end{equation}
We choose frequencies $\omega_1 = 1.3$ and $\omega_2 = 8.48$ and draw the elements of $\wvec$ independently from a uniform distribution on the open interval $(0,1)$.
The vectors $\vvec_1$ and $\vvec_2$ are the oscillatory spatial modes that we want to recover using compressed sensing.
For illustrative purposes, we choose the Gaussians
\begin{equation}
    \begin{aligned}
        \vvec_1 &= 2 \exp\left(-\frac{(x-0.5)^2}{2 (0.6)^2} - \frac{(y-0.5)^2}{2 (0.2)^2}\right) \\
        \vvec_2 &= \exp\left(-\frac{(x+0.25)^2}{2 (0.6)^2} - \frac{(y-0.35)^2}{2 (1.2)^2}\right),
    \end{aligned}
    \label{eq:toy_ex_true_modes}
\end{equation}
where $x$ and $y$ are spatial coordinates.
Figure~\ref{fig:toy_ex_true_modes} shows a visualization of these modes.

\begin{figure}[b]
    \centering
    \includegraphics{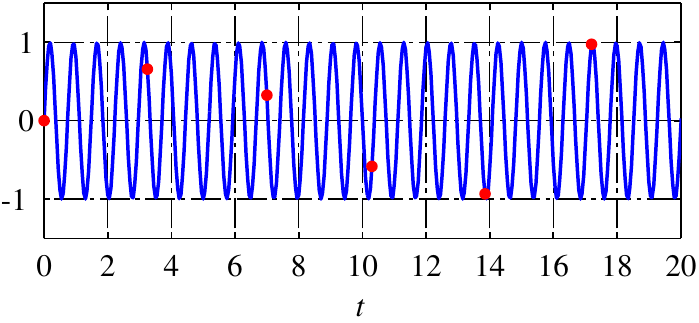}
    \caption{Random sampling for the canonical dataset.
    The first six sample points are plotted over a sine wave with frequency $\omega_2 = 8.48$, the fastest component of the signal defined by Eq.~\eqref{eq:canon_signal}.
    Clearly, the sample points do not resolve the fastest oscillations.}
    \label{fig:toy_ex_sampling}
\end{figure}

We generate a nominal signal $\Fmat$ whose columns are given by $\fvec(j\Delta t)$ for $j = 0,1,\ldots,n-1$, with $\Delta t = 0.05$ and $n = 8001$.
The signal is sampled with a minimum spacing $s_\text{min} = 60$ and a maximum spacing $s_\text{max} = 75$, resulting in 117 total samples.
Since the fastest frequency in the signal is $\omega_2 = 8.48$ and the underlying timestep is $\Delta t = 0.05$, the sample spacing that satisfies the Nyquist-Shannon sampling criterion is $s_\text{Nyq} = 7$.
Thus we see that we are at best sampling at eight times slower than required by the Nyquist-Shannon sampling criterion.
Figure~\ref{fig:toy_ex_sampling} shows a plot of $\sin(8.48t)$ overlaid with points corresponding to the random sampling.
It is clear that using traditional techniques, there is not enough data to reconstruct the underlying signal.

\begin{figure}[b]
    \centering
    \includegraphics{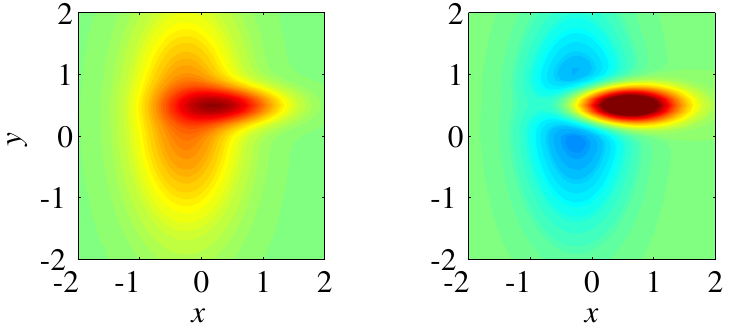}
    \caption{POD modes of the canonical dataset.
    The POD modes mix together features of the true oscillatory modes (Fig.~\ref{fig:toy_ex_true_modes}).}
    \label{fig:toy_ex_pod_modes}
\end{figure}

By construction, this signal is compressible, consisting of two dominant oscillations and low-amplitude broadband noise.
It is thus suitable for compressed sensing.
Furthermore, we see from Fig.~\ref{fig:toy_ex_pod_modes} that POD modes computed from $F$ are not aligned with the oscillatory ones.\footnote{We find that the POD modes do not differ much when computed from the time-resolved signal $F$ versus the sampled signal.}
Rather, each POD mode combines features of both oscillatory modes.
This is by design; for our method to work properly, it must correctly combine the POD modes such that these features are correctly isolated.

\begin{figure}[b]
    \centering
    \includegraphics{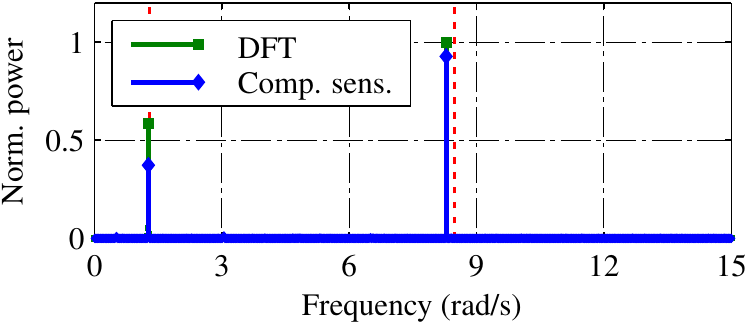}
    \caption{Comparison of spectra for the canonical dataset.
    Power values are normalized by the peak power computed using a DFT of the time-resolved data.
    The compressed sensing computation very accurately identifies the expected frequencies.
    (The true frequencies are denoted by red dotted lines.)}
    \label{fig:toy_ex_spectrum}
\end{figure}

Figure~\ref{fig:toy_ex_spectrum} shows the result of solving Eq.~\eqref{eq:cs_mat_opt_problem_POD} using the software package \texttt{cvx}~\citep{grantGINCP08,grantCVXURL12} to compute $\Ahatmat_r$.
We see that using compressed sensing, we correctly identify the two dominant frequencies, with less than 2.5~\% error in each case.
In fact, the identified frequencies agree exactly with those computed from a DFT of the time-resolved data, and are thus as accurate as can be expected.
With regard to the spectral power values, we find good agreement for the primary frequency, but noticeable error for the secondary frequency.
The rest of the frequencies have neglibible energy, a result of the $\ell_1$ minimization.

\begin{figure}[t]
    \centering
    \includegraphics{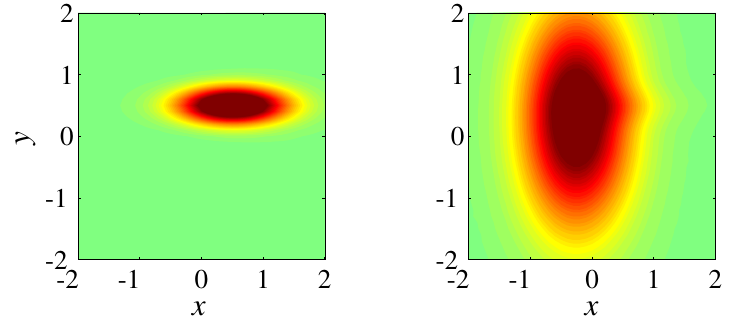}
    \caption{Compressed sensing modes for the canonical dataset.
    Comparing to Figs.~\ref{fig:toy_ex_true_modes} and~\ref{fig:toy_ex_pod_modes}, it is clear that compressed sensing has correctly combined the POD modes to recover the original oscillatory modes.}
    \label{fig:toy_ex_cs_modes}
\end{figure}

From Fig.~\ref{fig:toy_ex_cs_modes}, we can see that the correct oscillatory modes are identified.
There are some aberrations, but for the most part the compressed sensing modes resemble the Gaussians shown in Fig.~\ref{fig:toy_ex_true_modes}, rather than the POD modes shown in Fig.~\ref{fig:toy_ex_pod_modes}.
We note that because this method relies on random sampling, if we repeat the computation, the aberrations are sometimes larger or smaller.
However, we can decrease the likelihood of such errors by simply taking more samples (either by sampling faster or by using a longer signal); this also decreases the error in the spectral power values.
Overall, Figs.~\ref{fig:toy_ex_spectrum} and~\ref{fig:toy_ex_cs_modes} show that our method is capable of identifying temporally oscillating structures in a spatial signal using sub-Nyquist-rate data, even in the presence of noise.

\subsection{Flow past a cylinder}

The low-Reynolds number flow past a cylinder leads to sustained oscillations in the wake.
The resulting wake structures are known collectively as a von K\'{a}rm\'{a}n vortex street.
It is well known that a von K\'{a}rm\'{a}n vortex street is dominated by a single characteristic frequency.
Thus while the flow may not be exactly sparse (in frequency space), it is an example of the type of flow that one might want to investigate experimentally using compressed sensing techniques.
As such, it provides a valuable test of our method.

\begin{figure}[b]
    \centering
    \includegraphics[width=3.3in]{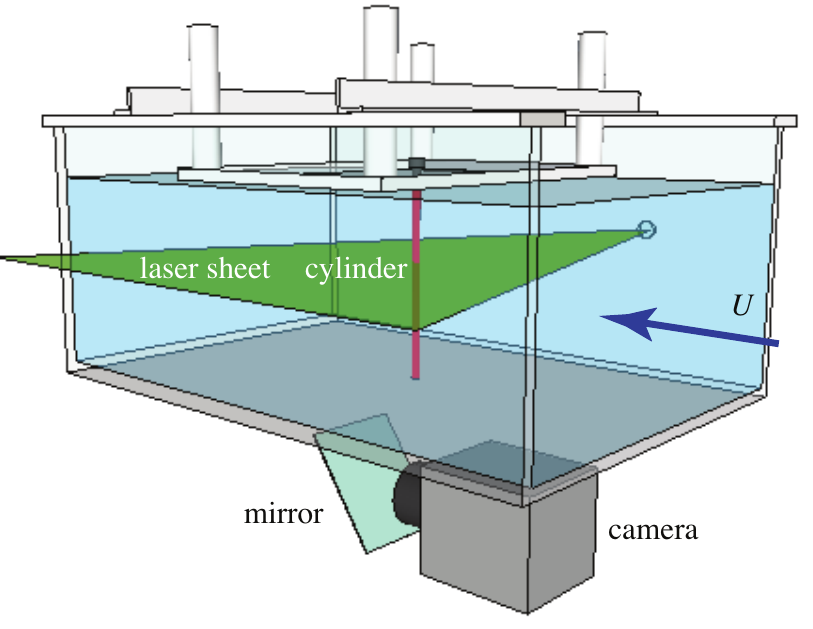}
    \caption{Experimental configuration for acquiring PIV data in the flow past a cylinder.
    The cylinder (red) is mounted vertically, and the wake is imaged in a horizontal plane located at the mid-span (green).}
    \label{fig:cyl_exp_schematic}
\end{figure}

We conduct a cylinder flow experiment in a recirculating, free-surface water channel.
The cylinder is made of anodized aluminum and has diameter $D = 9.5$~mm and length $L = 260$~mm.
With a freestream velocity $U_\infty = 4.35$~cm/s, this yields a Reynolds number $\text{Re} = 413$ (based on the cylinder diameter).
To eliminate the effect of surface waves, we suspend the cylinder vertically in the test section using an acrylic plate placed over the upper boundary of the water channel, as shown in Fig.~\ref{fig:cyl_exp_schematic}.
The gap between the cylinder and the lower wall of the channel is kept small (1--2~mm) to minimize three-dimensional effects.

We generate a laser sheet using a Nd:YAG laser (Litron Nano L 50-50) and illuminate the cross-section at the mid-span of the cylinder.
The sheet is imaged from below the water channel with a hybrid CCD/CMOS camera (LaVision, Imager sCMOS).
The laser and the camera are synchronized with a programmable timing unit.
We acquire 8000 image pairs with a delay of 8000~$\mu$s between exposures, at an overall sampling frequency of 20~Hz.
For seeding, we use neutrally buoyant hollow ceramic spheres with an average diameter of 10~$\mu$m.

PIV velocity fields are computed with a spatial cross-correlation algorithm using LaVision DaVis 8.1.2 software.
The data are processed using four passes with 50~\% overlap: one pass with a $128 \times 128$~pixel interrogation window, one pass with a $64 \times 64$~pixel window, and two passes with a $32 \times 32$ pixel window.
This results in velocity fields with $2160 \times 1280$~pixel resolution, with a cylinder diameter of approximately 128~pixels.
The final vector fields have a resolution of $135 \times 80$.

\begin{figure}[b]
    \centering
    \includegraphics{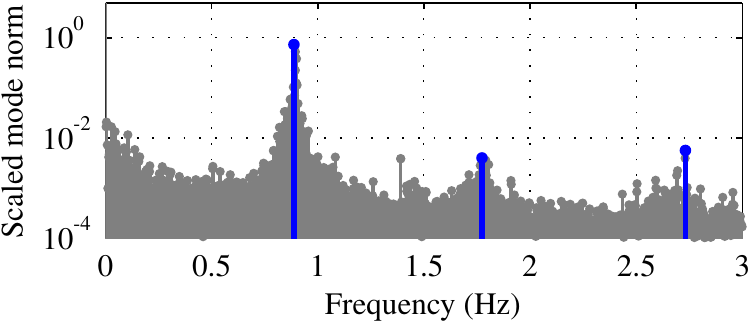}
    \caption{DMD spectrum computed from the flow past a cylinder using time-resolved PIV data.
    (The mode norms are scaled as described in~\cite{tuArxiv13}, in order to eliminate spurious peaks.)
    Peaks corresponding to modes shown in Fig.~\ref{fig:cyl_exp_dmd_modes} are highlighted in blue.
    There is a dominant spectral peak at $f = 0.889$~Hz, corresponding to the wake shedding frequency.
    Superharmonics of this frequency also appear in the spectrum, but with much lower energy.}
    \label{fig:cyl_exp_dmd_spectrum}
\end{figure}

For $\text{Re} = 413$, a sampling rate of 20~Hz easily resolves the wake shedding frequency, which can be estimated to be on the order of 1~Hz.
Thus we can use the time-resolved PIV data to compute DMD modes and eigenvalues.
These provide a basis of comparison for our method, as they are in effect the true oscillatory modes and frequencies that we are trying to approximate using compressed sensing.
The resulting DMD spectrum is shown in Fig.~\ref{fig:cyl_exp_dmd_spectrum}.
(To eliminate spurious peaks, the mode norms are scaled as described in~\cite{tuArxiv13}.)
We observe that there is a dominant frequency at $f_\text{wake} = 0.889$.
There are also harmonic peaks in the spectrum at approximately $2 f_\text{wake}$ and $3 f_\text{wake}$.

\begin{figure*}
    \centering
    \includegraphics{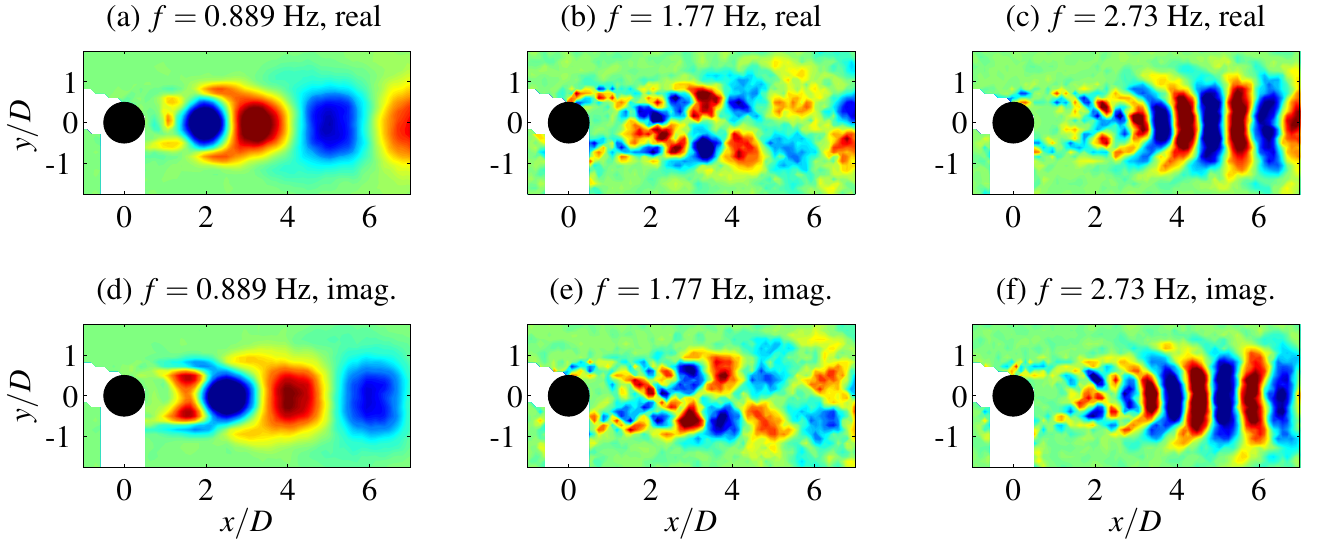}
    \caption{DMD modes computed from the flow past a cylinder, illustrated using contours of vorticity.
    The figures in the top row show the real part of each mode; the imaginary parts are shown in the bottom row.}
    \label{fig:cyl_exp_dmd_modes}
\end{figure*}

We note that the peaks in the spectrum are somewhat broad, and that the superharmonic peaks are significantly lower than the peak corresponding to the shedding frequency.
Thus while we can identify three spectral peaks, one could argue that the flow is in fact dominated by a single frequency.
The DMD modes corresponding to the wake frequency show strong coherence and top-bottom symmetry (Fig.~\ref{fig:cyl_exp_dmd_modes}~(a)), as do those corresponding to $3 f_\text{wake}$ (Fig.~\ref{fig:cyl_exp_dmd_modes}~(c)).
The modes corresponding to $2 f_\text{wake}$ show features of top-bottom anti-symmetry, but the structures are less coherent.

\begin{figure*}
    \centering
    \includegraphics{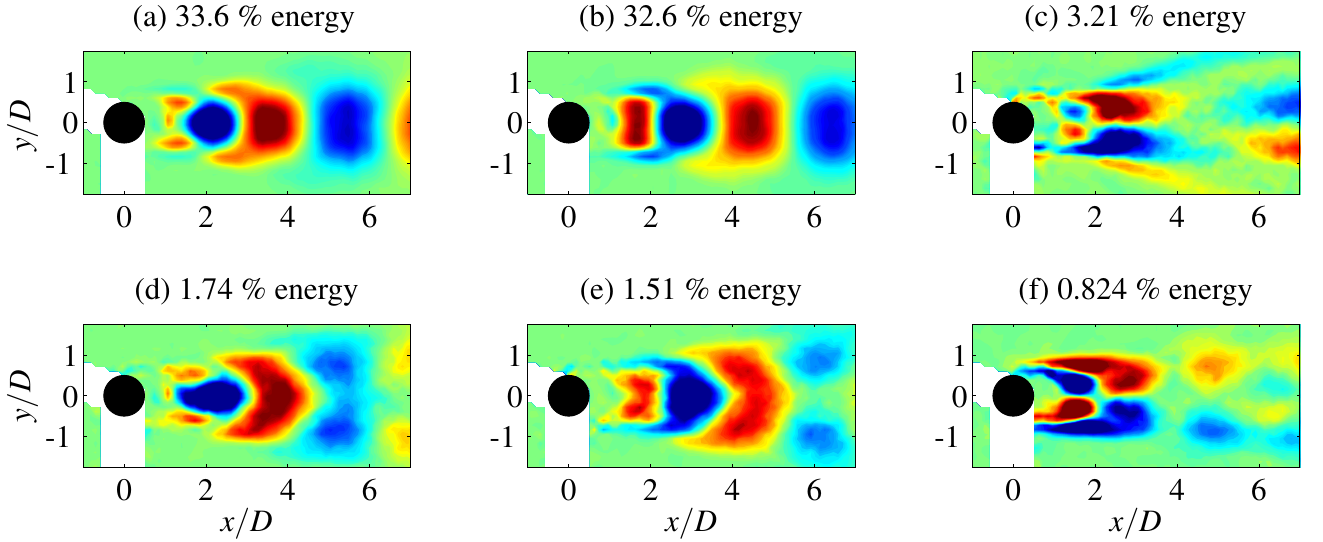}
    \caption{First six POD modes computed from the flow past a cylinder, illustrated using contours of vorticity.
    The dominant two modes ((a)~and(b)) resemble the DMD modes corresponding to the wake shedding frequency (Figs.~\ref{fig:cyl_exp_dmd_modes}~(a) and~(d)).
    The next four modes contain coherent structures, but do not resemble DMD modes.}
    \label{fig:cyl_exp_pod_modes}
\end{figure*}

\begin{figure*}
    \centering
    \includegraphics{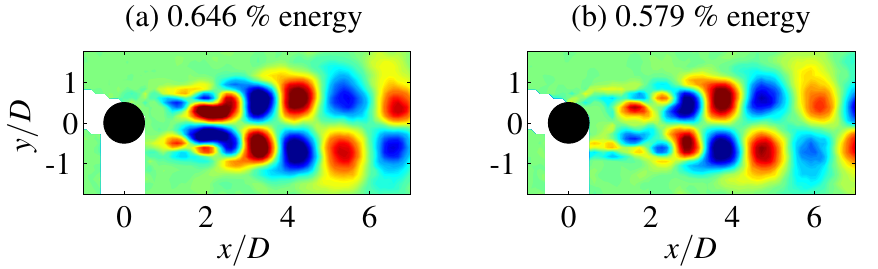}
    \caption{DMD-like POD modes computed from the flow past a cylinder, illustrated using contours of vorticity.
    These two modes resemble the higher-frequency DMD modes shown in Figs.~\ref{fig:cyl_exp_dmd_modes}~(b) and~(e).
    However, they contain very little energy.
    (Left) Mode index 7, 0.646~\% energy; (right) Mode index 8, 0.579~\% energy.}
    \label{fig:cyl_exp_pod_modes_2}
\end{figure*}

\begin{figure*}
    \centering
    \includegraphics{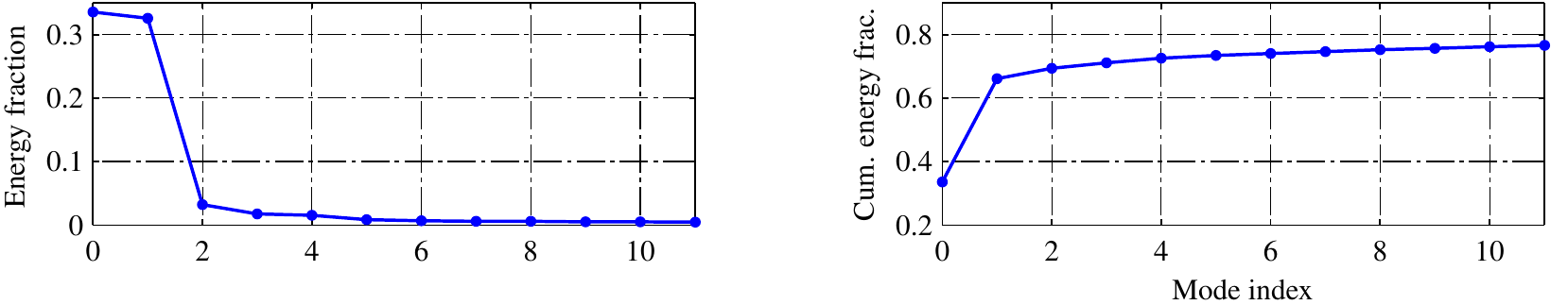}
    \caption{POD energy content for the flow past a cylinder.
    The first two modes dominate, followed by a slow roll-off in energy content.
    12 modes are required to capture 75~\% of the energy in the dataset.
    (Left) Energy fraction per mode; (right) cumulative energy fraction.}
    \label{fig:cyl_exp_pod_energy}
\end{figure*}

Figure~\ref{fig:cyl_exp_pod_modes} shows the six most energetic POD modes, computed using time-resolved PIV data.\footnote{In practice, we would compute the POD modes using only the sampled data, and not the time-resolved data.
However, for this flow we do not expect the POD basis to change much if computed from the sampled data, due to the strong attraction of the dynamics onto the low-dimensional POD subspace.}
Comparing Figs.~\ref{fig:cyl_exp_pod_modes}~(a) and~(b) to Figs.~\ref{fig:cyl_exp_dmd_modes}~(a) and~(d), we see that the first two POD modes resemble the DMD modes corresponding to the wake shedding frequency.
The remainder of the first six POD modes contain coherent structures, but do not resemble DMD modes.
However, if we consider even lower energy modes, we do find some that resemble higher-frequency DMD modes; these POD modes are shown in Fig.~\ref{fig:cyl_exp_pod_modes_2}.
The POD energy distribution is illustrated in Fig.~\ref{fig:cyl_exp_pod_energy}.
We see that the flow is dominated by a single pair of POD modes.
There is a sharp drop-off in energy content thereafter, with 12 modes required to capture 75~\% of the energy contained in the dataset.
We choose these first 12 modes as our low-dimensional basis for compressed sensing (see Sec.~\ref{ssec:pod-projection}).

\begin{figure}
    \centering
    \includegraphics{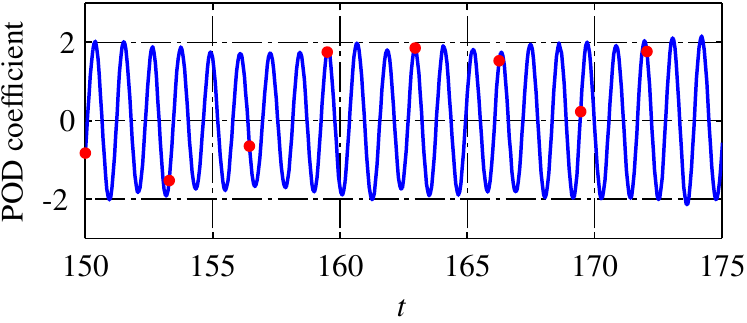}
    \caption{Random sampling for the flow past a cylinder.
    Eight consecutive sample points are plotted over the time history of the first POD coefficient.
    We see that the sample points clearly do not resolve the fastest oscillations in the signal.}
    \label{fig:cyl_exp_sampling}
\end{figure}

For the compressed sensing computation, we downsample the time-resolved PIV data, rather than acquiring a new dataset using a trigger.
We sample the data using the minimum/maximum separation strategy, choosing $s_\text{min} = 50$ and $s_\text{max} = 70$, in comparison to $s_\text{Nyq} = 10$.
This results in 33 total samples, out of the original $n = 2000$.
Due to the similarity of the first POD mode and the dominant DMD mode, we expect that a time history the first POD coefficient will contain oscillations at the wake shedding frequency.
Fig.~\ref{fig:cyl_exp_sampling} shows the sample points overlaid on a time trace of the first POD coefficient.
We see that again, the sample points are so infrequent that traditional methods would not be able to reconstruct the underlying signal.

\begin{figure}
    \centering
    \includegraphics{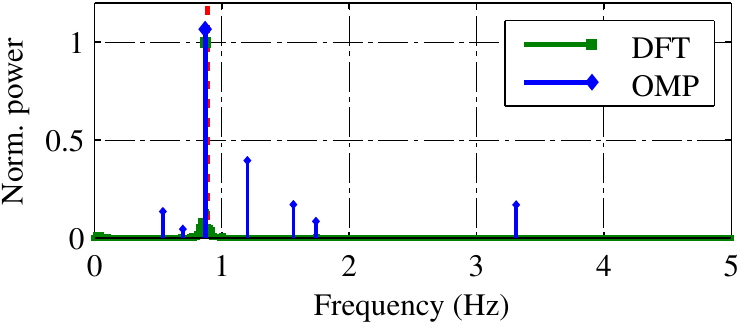}
    \caption{Comparison of spectra for the flow past a cylinder.
    Power values are normalized by the peak power computed using a DFT of the time-resolved data.
    OMP correctly identifies the wake shedding frequency.
    (The true frequency is denoted by a red dotted line.)}
    \label{fig:cyl_exp_omp_spectrum}
\end{figure}

For this larger computation, we perform compressed sensing using OMP rather than $\ell_1$ minimization.
We compute the first ten DFT modes using this greedy approach; the resulting spectrum is shown in Fig.~\ref{fig:cyl_exp_omp_spectrum}.
Once again, a dominant peak is identified, here corresponding to the wake shedding frequency.
The identified frequency again agrees exactly with that computed from a DFT of the time-resolved data.
The error with respect to the true wake frequency is less than 2.5~\%.
There is also good agreement between the OMP and DFT results in terms of the energy associated with the dominant frequency.

Unfortunately, the harmonic peaks observed in the DMD spectrum (Fig.~\ref{fig:cyl_exp_dmd_spectrum}) do not appear here.
This is the case even as we vary the sampling rate and the total number of samples.
(As we do so, the dominant peak is consistently identified, but there is no pattern in the other peaks that appear.)
We see from Fig.~\ref{fig:cyl_exp_omp_spectrum} that a DFT computed from the time-resolved POD coefficients does not identify harmonic peaks either.
Since we assume our signal is sparse in the Fourier basis (see Sec.~\ref{ssec:basis_choice}), the best we can expect of our OMP computation is agreement with a DFT.
Thus it should be expected that OMP only identifies one dominant peak; the rest are spurious, explaining why they vary in a seemingly random way as the computation is repeated with different random samples.

It is interesting that for this dataset, DMD identifies harmonic spectral peaks while a DFT identifies only the fundamental frequency.
While an in-depth comparison of DMD and Fourier analyses lies outside the scope of this work, we highlight a few key points.
First, unlike DFT modes, DMD modes are not orthogonal.
This is why the DMD spectrum shown in Fig.~\ref{fig:cyl_exp_dmd_spectrum} cannot be interpreted as a plot of power spectral density.
(As such, we do not expect the magnitudes of the peaks in Figs.~\ref{fig:cyl_exp_dmd_spectrum} and~\ref{fig:cyl_exp_omp_spectrum} to agree.)
Second, the temporal evolution associated with each DMD mode may include exponential growth or decay, in contrast to the purely oscillatory dynamics of DFT modes.
It is likely due to these differences that the DMD and DFT spectra disagree regarding the frequency content of the signal.

\begin{figure}
    \centering
    \includegraphics{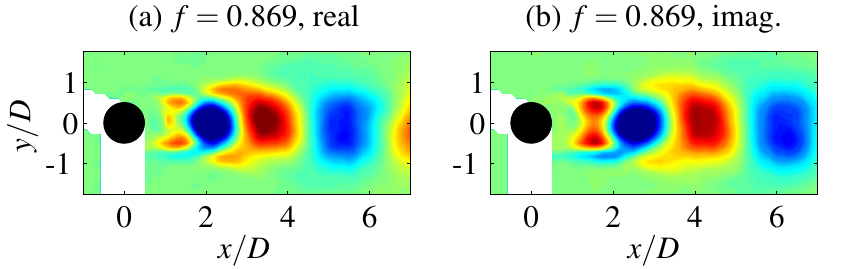}
    \caption{Dominant OMP mode for the flow past a cylinder, illustrated using contours of vorticity.
    Comparing to Figs.~\ref{fig:cyl_exp_dmd_modes}~(a) and~(b), we see that OMP identifies the general structure of the true oscillatory mode.
    (Left) Real part; (right) imaginary part.}
    \label{fig:cyl_exp_omp_modes}
\end{figure}

The dominant OMP mode is depicted in Fig.~\ref{fig:cyl_exp_omp_modes}.
We see that OMP correctly pairs the dominant POD modes with the wake shedding frequency, as expected based on DMD analysis.
(Recall that the dominant POD modes closely resemble the dominant DMD modes.)
Though the OMP modes do not exactly match the DMD modes (Figs.~\ref{fig:cyl_exp_dmd_modes}~(a) and~(d)), they capture the main coherent structures.
We note that in theory, one could compute the POD modes using non-time-resolved data and then independently measure the dominant flow frequency using a hot wire (which is much faster than PIV and can resolve the wake shedding frequency).
One could then pair these together to arrive at the same conclusions as we get using OMP.
However, the compressed sensing/OMP approach identifies the oscillatory modes and corresponding frequency directly from the data and does not require a priori knowledge of the flow dynamics (aside from an intuition that the signal is compressible).
As such, it is generalizable to more complex flows, where it may not be obvious how to pair the dominant POD modes with characteristic flow frequencies.


\section{Conclusions and future work}
\label{sec:conclusions}

We have developed a method for computing temporally oscillating spatial modes from sub-Nyquist-rate data.
This method combines concepts from compressed sensing, formulated for vector-valued signals, and DMD.
The result is a method that is similar to the DFT, as the modes each correspond to a particular temporal oscillation frequency, but no temporal growth or decay.
An interesting future direction would be to relax this assumption, either by choosing a more general basis $\Psimat$ or by adaptively searching for the correct growth/decay rates.
Doing so would yield a method that could truly be called ``compressed DMD.''

We demonstrated the capabilities of our method using both numerical and experimental data.
The compressed sensing computations were done using $\ell_1$ minimization and greedy algorithms (specifically OMP), respectively.
In both cases, the correct frequencies and spatial modes were identified.
This verifies not only the compressed sensing approach, but also the random sampling strategy proposed in Sec.~\ref{ssec:sampling-strategy}.

While our method was originally motivated by the limitations of PIV technology, the strategies we employed apply more generally, as they deal with two challenges that commonly arise in the data-driven analysis of complex dynamical systems.
The first of these is insufficient temporal resolution.
Using compressed sensing, we can extend the capabilities of sensors that are limited by slow sampling rates.
This approach is also beneficial with faster sensors, for which compressed sensing can be leveraged to analyze even faster dynamics.

In addition, our method successfully deals with so-called ``big data'' problems.
Even when fast sensors are available, uniform sampling can lead to an overabundance of data.
The use of compressed sensing can be thought of as method for more efficient sampling, reducing data storage requirements.
However, even if temporal sampling is reduced, large measurement vectors can strain computational resources.
This is the case for PIV, due to the desire for increased spatial/image resolution.
As a result, processing PIV data can be cumbersome.
To overcome this, our method uses POD to encode PIV data in a low-dimensional subspace, making the data amenable for standard compressed sensing algorithms.

Given the broad applicability of compressed sensing and dimensionality reduction, it is likely that our method will be applicable to a wide-range of dynamical systems.
Any system limited by sampling rates, storage capacity, or high-dimensional measurement vectors would be a prime target.
In its current form, our method enables spectral analysis when standard approaches may fail.
Future extensions could explore the use of bases other than the Fourier basis, providing more general characterizations of a system.
Certainly, the continued merging of compressed sensing and dynamical systems concepts holds much promise.


\begin{acknowledgements}
    The authors acknowledge funding from the AFOSR and the NSF, and thank Steve Brunton for many insightful discussions regarding compressed sensing in the context of dynamical systems.
\end{acknowledgements}


\bibliographystyle{spbasic}
\bibliography{jabbrv,references}

\begin{thebibliography}{41}
\providecommand{\natexlab}[1]{#1}
\providecommand{\url}[1]{{#1}}
\providecommand{\urlprefix}{URL }
\expandafter\ifx\csname urlstyle\endcsname\relax
  \providecommand{\doi}[1]{DOI~\discretionary{}{}{}#1}\else
  \providecommand{\doi}{DOI~\discretionary{}{}{}\begingroup
  \urlstyle{rm}\Url}\fi
\providecommand{\eprint}[2][]{\url{#2}}

\bibitem[{Bai et~al.(2013)Bai, Wimalajeewa, Berger, Wang, Glauser, and
  Varshney}]{baiAIAA13}
Bai Z, Wimalajeewa T, Berger Z, Wang G, Glauser MN, Varshney PK (2013) Physics
  based compressive sensing approach applied to airfoil data collection and
  analysis. AIAA Paper 2013-0772, 51st AIAA Aerospace Sciences Meeting and
  Exhibit

\bibitem[{Baraniuk(2007)}]{baraniukIEEESPM07}
Baraniuk RG (2007) Compressive sensing. IEEE Signal Proc\ Mag 24(4):118--124

\bibitem[{Brunton et~al.(2013{\natexlab{a}})Brunton, Brunton, Proctor, and
  Kutz}]{bruntonbArxiv13}
Brunton BW, Brunton SL, Proctor JL, Kutz JN (2013{\natexlab{a}}) Optimal sensor
  placement and enhanced sparsity for classification. arXiv:1310.4217 [cs.CV]

\bibitem[{Brunton et~al.(2013{\natexlab{b}})Brunton, Proctor, and
  Kutz}]{bruntonsArxiv13b}
Brunton SL, Proctor JL, Kutz JN (2013{\natexlab{b}}) Compressive sampling and
  dynamic mode decomposition. arXiv:1312.5186v1 [math.DS]

\bibitem[{Brunton et~al.(2013{\natexlab{c}})Brunton, Tu, Bright, and
  Kutz}]{bruntonsArxiv13}
Brunton SL, Tu JH, Bright I, Kutz JN (2013{\natexlab{c}}) Compressive sensing
  and low-rank libraries for classification of bifurcation regimes in nonlinear
  dynamical systems. arXiv:1312.4221v1 [math.DS]

\bibitem[{Bryan and Leise(2013)}]{bryanSIAMR13}
Bryan K, Leise T (2013) Making do with less: an introduction to compressed
  sensing. SIAM Rev 55(3):547--566

\bibitem[{Candes(2008)}]{candesCRASP08}
Candes EJ (2008) The restricted isometry property and its implications for
  compressed sensing. C R Acad\ Sci\ Paris S{\'e}r~I Math 346(9--10):589--592

\bibitem[{Candes and Tao(2005)}]{candesIEEETIT05}
Candes EJ, Tao T (2005) Decoding by linear programming. IEEE T Inform\ Theory
  51(12):4203--4215

\bibitem[{Candes and Tao(2006)}]{candesIEEETIT06_2}
Candes EJ, Tao T (2006) Near-optimal signal recovery from random projections:
  Universal encoding strategies? IEEE T Inform\ Theory 52(12):5406--5425

\bibitem[{Candes and Wakin(2008)}]{candesIEEESPM08}
Candes EJ, Wakin MB (2008) An introduction to compressive sampling. IEEE Signal
  Proc\ Mag 25(2):21--30

\bibitem[{Candes et~al.(2006)Candes, Romberg, and Tao}]{candesIEEETIT06_1}
Candes EJ, Romberg J, Tao T (2006) Robust uncertainty principles: Exact signal
  reconstruction from highly incomplete frequency information. IEEE T Inform\
  Theory 52(2):489--509

\bibitem[{Chen and Huo(2006)}]{chenIEEETSP06}
Chen J, Huo X (2006) Theoretical results on sparse representations of
  multiple-measurement vectors. IEEE T Signal Proces 54(12):4634--4643

\bibitem[{Cotter et~al.(2005)Cotter, Rao, Engan, and
  Kreutz-Delgado}]{cotterIEEETSP05}
Cotter SF, Rao BD, Engan K, Kreutz-Delgado K (2005) Sparse solutions to linear
  inverse problems with multiple measurement vectors. IEEE T Signal Proces
  53(7):2477--2488

\bibitem[{Dick et~al.(2000)Dick, Harris, and Rice}]{dickIEEESFPCCM00}
Dick C, Harris F, Rice M (2000) Synchronization in software radios --- carrier
  and timing recovery using fpgas. In: Proceedings of the 2000 IEEE Symposium
  on Field-Programmable Custom Computing Machines, IEEE, pp 195--204

\bibitem[{Donoho(2006)}]{donohoIEEETIT06b}
Donoho DL (2006) Compressed sensing. IEEE T Inform\ Theory 52(4):1289--1306

\bibitem[{Duarte et~al.(2008)Duarte, Davenport, Takhar, Laska, Sun, Kelly, and
  Baraniuk}]{duarteIEEESPM08}
Duarte MF, Davenport MA, Takhar D, Laska JN, Sun T, Kelly KF, Baraniuk RG
  (2008) Single-pixel imaging via compressive sampling. IEEE Signal Proc\ Mag
  25(2):83--91

\bibitem[{Eldar and Mishali(2009)}]{eldarIEEETIT09}
Eldar YC, Mishali M (2009) Robust recovery of signals from a structured union
  of subspaces. IEEE T Inform\ Theory 55(11):5302--5316

\bibitem[{Gamper et~al.(2008)Gamper, Boesiger, and Kozerke}]{gamperMRM08}
Gamper U, Boesiger P, Kozerke S (2008) Compressed sensing in dynamic {MRI}.
  Magn\ Reson\ Med 59(2):365--373

\bibitem[{Grant and Boyd(2008)}]{grantGINCP08}
Grant M, Boyd SP (2008) Graph implementations for nonsmooth convex programs.
  In: Blondel V, Boyd SP, Kimura H (eds) Recent Advances in Learning and
  Control, Lecture Notes in Control and Information Sciences, vol 371,
  Springer-Verlag, pp 95--110

\bibitem[{Grant and Boyd(2012)}]{grantCVXURL12}
Grant M, Boyd SP (2012) {CVX}: {M}atlab software for disciplined convex
  programming, version 2.0 beta. \url{http://cvxr.com/cvx},
  \urlprefix\url{http://cvxr.com/cvx}

\bibitem[{Herman and Strohmer(2009)}]{hermanIEEETSP09}
Herman MA, Strohmer T (2009) High-resolution radar via compressed sensing. IEEE
  T Signal Proces 57(6):2275--2284

\bibitem[{Jovanovi\'c et~al.(2013)Jovanovi\'c, Schmid, and
  Nichols}]{jovanovicArxiv13}
Jovanovi\'c MR, Schmid PJ, Nichols JW (2013) Sparsity-promoting dynamic mode
  decomposition. arXiv:1309.4165v1 [physics.flu-dyn]

\bibitem[{Lustig et~al.(2007)Lustig, Donoho, and Pauly}]{lustigMRM07}
Lustig M, Donoho DL, Pauly JM (2007) Sparse {MRI}: the application of
  compressed sensing for rapid {MR} imaging. Magn\ Reson\ Med 58(6):1182--1195

\bibitem[{Malioutov et~al.(2005)Malioutov, \c{C}etin, and
  Willsky}]{malioutovIEEETSP05}
Malioutov D, \c{C}etin M, Willsky AS (2005) A sparse signal reconstruction
  perspective for source localization with sensor arrays. IEEE T Signal Proces
  53(8, 2):3010--3022

\bibitem[{Needell and Tropp(2009)}]{needellACHA09}
Needell D, Tropp JA (2009) {CoSaMP}: Iterative signal recovery from incomplete
  and inaccurate samples. Appl\ Comput\ Harmon\ A 26(3):301--321

\bibitem[{Nyquist(1928)}]{nyquistTAIEE28}
Nyquist H (1928) Certain topics in telegraph transmission theory. T AIEE
  47(2):617--644

\bibitem[{Potter et~al.(2010)Potter, Ertin, Parker, and
  \c{C}etin}]{potterPIEEE10}
Potter LC, Ertin E, Parker JT, \c{C}etin M (2010) Sparsity and compressed
  sensing in radar imaging. P IEEE 98(6):1006--1020

\bibitem[{Romberg(2008)}]{rombergIEEESPM08}
Romberg J (2008) Imaging via compressive sampling. IEEE Signal Proc\ Mag
  25(2):14--20

\bibitem[{Rowley(2005)}]{rowleyIJBC05}
Rowley CW (2005) Model reduction for fluids, using balanced proper orthogonal
  decomposition. Int\ J Bifurcat\ Chaos 15(3):997--1013

\bibitem[{Rowley et~al.(2009)Rowley, Mezi\'{c}, Bagheri, Schlatter, and
  Henningson}]{rowleyJFM09}
Rowley CW, Mezi\'{c} I, Bagheri S, Schlatter P, Henningson DS (2009) Spectral
  analysis of nonlinear flows. J Fluid Mech 641:115--127

\bibitem[{Schmid(2010)}]{schmidJFM10}
Schmid PJ (2010) Dynamic mode decomposition of numerical and experimental data.
  J Fluid Mech 656:5--28

\bibitem[{Shannon(1949)}]{shannonPIRE49}
Shannon CE (1949) Communication in the presence of noise. P IRE 37(1):10--21

\bibitem[{Sirovich(1987)}]{sirovichQAM87_2}
Sirovich L (1987) Turbulence and the dynamics of coherent structures part {II}:
  {S}ymmetries and transformations. Q\ Appl\ Math 45(3):573--582

\bibitem[{Taubman and Marcellin(2001)}]{taubmanJPEG2000ICFSP01}
Taubman DS, Marcellin MW (2001) {JPEG} 2000: Image Compression Fundamentals,
  Standards and Practice. Kluwer

\bibitem[{Tropp(2004)}]{troppIEEETIT04}
Tropp JA (2004) Greed is good: Algorithmic results for sparse approximation.
  IEEE T Inform\ Theory 50(10):2231--2242

\bibitem[{Tropp(2006)}]{troppSP06b}
Tropp JA (2006) Algorithms for simultaneous sparse approximation. {P}art {II}:
  Convex relaxation. Signal Process 86(3):589--602

\bibitem[{Tropp and Gilbert(2007)}]{troppIEEETIT07}
Tropp JA, Gilbert AC (2007) Signal recovery from random measurements via
  orthogonal matching pursuit. IEEE T Inform\ Theory 53(12):4655--4666

\bibitem[{Tropp et~al.(2006)Tropp, Gilbert, and Strauss}]{troppSP06a}
Tropp JA, Gilbert AC, Strauss MJ (2006) Algorithms for simultaneous sparse
  approximation. {P}art {I}: Greedy pursuit. Signal Process 86(3):572--588

\bibitem[{Tu et~al.(2013{\natexlab{a}})Tu, Griffin, Hart, Rowley,
  Cattafesta~III, and Ukeiley}]{tuEF13}
Tu JH, Griffin J, Hart A, Rowley CW, Cattafesta~III LN, Ukeiley LS
  (2013{\natexlab{a}}) Integration of non-time-resolved {PIV} and time-resolved
  velocity point sensors for dynamic estimation of velocity fields. Exp\ Fluids
  54(2)

\bibitem[{Tu et~al.(2013{\natexlab{b}})Tu, Rowley, Luchtenburg, Brunton, and
  Kutz}]{tuArxiv13}
Tu JH, Rowley CW, Luchtenburg DM, Brunton SL, Kutz JN (2013{\natexlab{b}}) On
  dynamic mode decomposition: Theory and applications. arXiv:1312.0041v1
  [math.NA]

\bibitem[{Wright et~al.(2009)Wright, Yang, Ganesh, Sastry, and
  Ma}]{wrightIEEETPAMI09}
Wright J, Yang AY, Ganesh A, Sastry SS, Ma Y (2009) Robust face recognition via
  sparse representation. IEEE T Pattern Anal 31(2):210--227

\end{thebibliography}

\end{document}